\documentclass[%
 reprint,
superscriptaddress,
 amsmath,amssymb,
 aps,
]{revtex4-1}

\newcommand{\mi}{{\mathrm i}}

\newcommand{\cmt}[2]{{[}#1,#2{]}}

\newcommand{\eas}[0]{\begin{eqnarray*}}
\newcommand{\eae}[0]{\end{eqnarray*}}
\newcommand{\les}[0]{\begin{equation}}
\newcommand{\lee}[0]{\end{equation}}
\newcommand{\leas}[0]{\begin{eqnarray}}
\newcommand{\leae}[0]{\end{eqnarray}}

\newcommand{\ed}{{\rm e}}
\newcommand{\bl}{{\rm b}}
\usepackage{ulem}
\usepackage{xcolor}
\newcommand{\cin}[1]{{\color{red}#1}}
\renewcommand{\cin}[1]{{#1}}
\newcommand{\cout}[1]{{\color{blue}#1}}
\renewcommand{\cout}[1]{}
\newcommand{\coutn}[1]{{\color{blue}#1}}
\renewcommand{\coutn}[1]{}
\newcommand{\crep}[2]{{\coutn{#1} \cin{#2}}}
\renewcommand{\sout}[1]{}

\newcommand{\cinf}[1]{{\color{blue}#1}}
\renewcommand{\cinf}[1]{{#1}}
\newcommand{\csoutf}[1]{{\color{blue}\sout{#1}}}
\renewcommand{\csoutf}[1]{}

\newcommand{\remove}[1]{{\color{blue}#1}}
\renewcommand{\remove}[1]{}
\usepackage{graphicx}
\usepackage{dcolumn}
\usepackage{bm}

\begin{document}

\preprint{APS/123-QED} 

\title{
Bulk-edge correspondence in  topological pumping}

\author{Y. Hatsugai}
\email{hatsugai.yasuhiro.ge@u.tsukuba.ac.jp}
\affiliation{Division of Physics, University of Tsukuba, Tsukuba 305-8571, Japan}
\author{T. Fukui}%
\email{takahiro.fukui.phys@vc.ibaraki.ac.jp}
\affiliation{Department of Physics, Ibaraki University, Mito 310-8512, Japan}


\date{\today}

\begin{abstract}
The topological pumping \cin{proposed in '80s and recently realized by cold atom experiments}\cite{Thouless83,Nakajima15,Lohse15} 
is revisited from
a view point of the bulk-edge correspondence.
For a system with boundaries,
a new form of the pumped charge is derived by the Berry connection
in the temporal gauge, that corresponds to the
shift  of  the center of mass (CM).
\coutn{ Shift of  the center of mass (CM) 
  as a pumped charge 
    is explicitly given by the 
      gauge invariant
    Berry connection
in the temporal gauge.}
\cin{Even with boundaries,
  the pumped charge is carried by the bulk and 
its quantization 
is guaranteed by 
the discontinuities of the CM associated with the edge states.
This is a modified Laughlin argument based on the local $U(1)$ invariance
although physics behind is quite different.
}
\cout{ A gauge invariant form of the pumped charge is proposed
 by the temporal gauge
 and is used to establish this bulk-edge correspondence.
 }
\end{abstract}

\pacs{Valid PACS appear here}
\maketitle


Fundamental stability of edge states \cite{Halperin82, Laughlin81, Hatsugai93-edge} in the quantum Hall
effect (QHE)
implies the phase
is topological.
The bulk is gapped but, with boundaries, there exists localized modes
in the gap.
Conversely non trivial topological phases are
necessarily associated with  edge states as generic localized modes
near the boundaries of the system or impurities.
It applies to any of the topological phases 
that include the fractional QH states,
quantum spin Hall states as the topological insulators, the
Haldane phase of integer spin chains and so on.
Generically emergence of the edge states
is reduced to the feature of the bulk,
that  is the bulk-edge correspondence \cite{Hatsugai93-bec}.

{\color{black} The phase is ``topological''
implies it is }``hidden''\cite{Wen89}, in the sense that
one can hardly observe its character in a bulk\cout{\cite{Wen89}},
even though the phase
is  characterized by a topological invariant\cite{Thouless82,KM05,Bernevig06}.
Further, \cout{in most cases} 
they are 
{mostly} not physical observables. Even in the QHE, whether  the observed
Hall conductance in 
experiments is of the bulk or of the edges is
still controversial.
The bulk topological character
is hidden but the edge states are real observables.
A typical example
is a surface Dirac cone of the three dimensional topological insulators,
that is observed by 
the angle resolved photoemission spectroscopies.
  \cout{clearly observe the Dirac cones
as the edge states for some of the semiconductors.}
The bulk-edge correspondence is thus
 a rigorous/conceptual bridge between the
 bulk and with boundaries,
 that was, at first, rigorously shown for the QHE \cite{Hatsugai93-bec} and
 justified today for the
others as well \cite{Kitaev01,Ryu02,KM05,Bernevig06,Qi06,Schulz00,Graf13}. 

This reverse way of thinking is \cout{crucially} important since one 
can realize why the edges are there from a
universal point of view.
Some of them are historically well-known such as
dangling bonds of semiconductors and some are new as
the chiral edge modes of photonic crystals \cite{Haldane08,Wang08} and
photon's \cite{Hafezi14}.
\cout{It is now clear that we do not need to restrict
ourselves to the quantum world. The photonic crystals are surely governed by
the classical Maxwell equation. }
This analogy even extends to the 
classical Newton equation
\cite{Prodan09,Kane14,Kariyado15}.
It implies the universal feature of the bulk-edge correspondence.

Ultimate developments of recent
quantum technology also trigger further studies of 
topological phases, such as realization of the Hofstadter butterfly
by a synthetic gauge field
in a artificial lattice using cold atoms \cite{Dalibard11}.
Many of the theoretical ideas for the topological phases and gauge structures
of matter 
are now directly measured experimentally where
quantum coherence and structures
are under the ultimate control and one can handle or
even synthesize 
dimensions \cite{Dalibard11,Fallani15}.
One of the most clear and fundamental examples is
a topological pump proposed long time ago \cite{Thouless83},
where the time is used as a synthetic dimension and the topological
effects in the QHE in 2D is realized in a simple one dimensional system.
Even though the proposal is quite old as the QHE,
it took so long time until its experimental realization 
  in cold atoms \cite{Nakajima15,Lohse15}, 
and intense studies have just began.
Although the bulk-edge correspondence is fundamental in
topological phases, it \cin{has never been} applied to this topological pump.
\cin{Here we have first made
  the role of the edge states clear in the topological pump.}
{\color{red}\sout{
We here revisit the problem and clarify
the bulk-edge correspondence in the
topological pump.
  Surprisingly it provides conceptually new aspects of the
  bulk-edge correspondence, that is again closely related to the experiments.}}
\cin{
  In this letter, a new expression of the pumped charge
is derived  
  by the Berry connection for the system with boundaries. 
  In the adiabatic limit, 
  the contribution of the the bulk and the edge are 
  clearly separated.
  The pumped charge is carried by the bulk but
  its quantization is guaranteed by the locality of the
  edge states. 
\csoutf{Although the topological pumping is redesciption of the
  quantum Hall effect in 2D, physical contents of the bulk-edge correspondence
  is quite different.} Local $U(1)$ gauge symmetry as the charge conservation
  is crucially important but plays a role \cinf{different from the QHE. 
Namely, in the QHE, the right- and left-edge states are always paired on the Fermi surface, 
whereas in the pumping, they can be
separately observed at different times.
Physically this is a fractionalization of the electron into the massive
Dirac fermions carrying half charge quantum.
}
}

Let us consider a many body
topological pump by
the time dependent one dimensional hamiltonian of
lattice free fermions of $L_x$ sites
\begin{eqnarray*}
  H(\theta,t )
  = \sum_{j}^{L_x}\big[ 
  -t_x^\theta c_{j+1}^\dagger c_j +h.c.
  + v_j(t) c_j ^\dagger c_j \big], 
\end{eqnarray*}
($t_x^\theta=t_x  e^{-\mi\theta/L_x}, t_x,t_y\in\mathbb{R}$. 
The twist $\theta$ is introduced to
define the current operator as well as the Berry connection 
 for a generic many-body state).
The time dependent potential $v_j(t)$ can be any {\color {black} if it is periodic as 
$v_{j+q}(t)=v_j(t)$ where $q$ is a positive integer.}
For simplicity, we choose as
$  v_j(t) =   -2t_y  \cos (  \frac {2\pi t}{T}- 2\pi \phi j)  $,
  $  \phi =p/q$
  with mutually prime $p$ and $q$. This is equivalent to the 2D QHE under
  a periodic potential\cite{Thouless82,Hatsugai93-edge,Hatsugai93-bec}.
  When mapping back to the QHE in 2D, the time as a synthetic dimension
  corresponds to
  the momentum in $y$ direction
  ($\frac {2\pi t}{T}\to k_y  $\cite{Hatsugai93-edge}).
  By writing the  one particle state as
  $|\psi \rangle  =\sum_j\psi_j c_j ^\dagger | 0 \rangle $, the wave function $\psi_j$
  satisfies the Harper equation,
  $-t_x^\theta \psi_{j-1\remove{,\ell_L}}-2 t_y \cos (2\pi\frac {t}{T}- 2\pi\phi j )\psi_{j\remove{,\ell_L}}-(t_x^\theta)^* \psi_{j+1\remove{,\ell_L}}=E_{\ell_L}\psi_{j\remove{,\ell_L}}$.
  The time-reversal (TR) is broken by a finite $\theta$ 
  but  is 
  recovered by taking the  $\theta\to 0$ limit after
  the calculation. The pumping period, $T$, controls the
  adiabaticity of the pumping.
  For simplicity, we take $T=2\pi$ and control the adiabaticity
  by the energy scale of $t_x$ and $t_y$.
  Two boundary conditions are discussed.
  One is the open boundary condition with edges, $\ed$: $\psi_0=\psi_{L_x}=0$ and the other is
  periodic one  for the bulk $\bl$: $\psi_{j+L_x}=\psi_j$.
  The many body eigen state of the snap shot hamiltonian $H(t)$ is given
  by specifying a set of occupied  one particle states $\{\ell_L\}$
  as
  $|\alpha \rangle =\prod_{\ell_L}(\sum_j\psi_{j,\ell_L}c_j ^\dagger |0 \rangle $.
  
\begin{figure}[h]
\includegraphics[width=80mm]{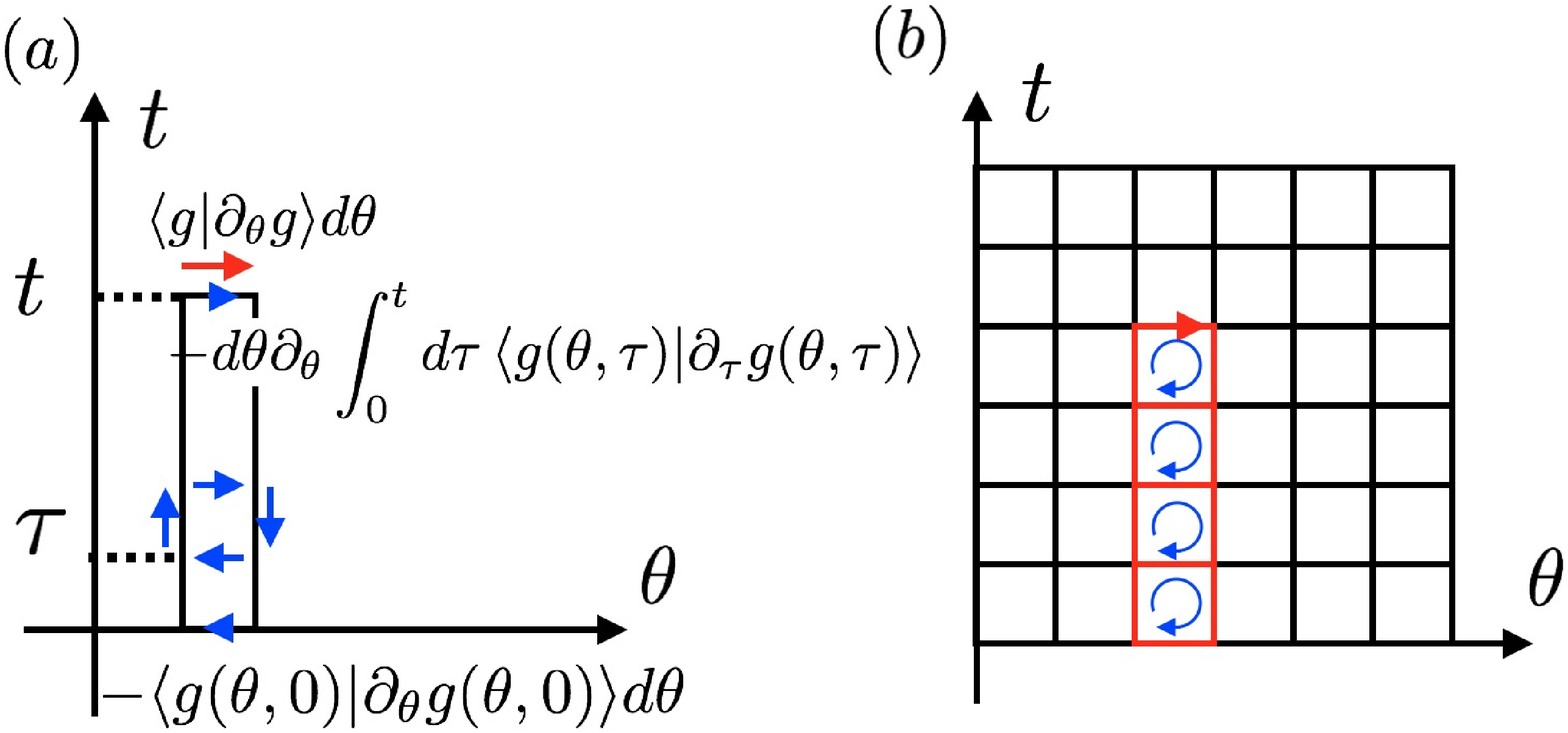}
\caption{\label{fig:t-gauge} 
  (a) 
  The Berry connection $A^{(t)}$ in the temporal gauge
  and (b) its lattice analogue \cite{Fukui05}.}
\end{figure}

  Using the current operator 
  \crep
      {
        $
        J
  = \frac {1}{L_x} (\mi \frac {t_x}{\hbar } e^{-\mi \theta/L_x})\sum_j  c_{j+1} ^\dagger c_j + h.c
= \hbar ^{-1} \partial_\theta H(\theta )
$,
  }
  {
  \begin{eqnarray*}
    J
&=&  \frac {1}{L_x} (\mi \frac {t_x}{\hbar } e^{-\mi \theta/L_x})\sum_j  c_{j+1} ^\dagger c_j + h.c
= \hbar ^{-1} \partial_\theta H(\theta,t ),
  \end{eqnarray*}
  }
  the measured current at the time $t$, 
  $  \delta j = \langle G(t)|J|G(t) \rangle - \langle g(t) | J| g(t) \rangle $,
  is evaluated by the adiabatic approximation \cite{Thouless83}
  assuming a finite energy gap above the snap shot
  ground state $|g(t) \rangle  $,
  $H(\theta,t )|g \rangle = | g \rangle E$. The state $|G(t) \rangle  $ is a
  true many body state 
  that obeys the time dependent Schr\"odinger equation, $\mi \hbar \partial _t |G(t) \rangle =H(\theta,t )|G(t) \rangle $. It reads
$
 \delta j 
  =
  -\mi B, \ B=
\partial _\theta A_t      -\partial _t A_\theta 
  $
  where $B$ is a field strength of the Berry connection $A_\mu = \langle g|\partial_\mu g \rangle $, $\mu =\theta,t $.
  Note that $ \langle g(t) | J| g(t) \rangle| _{\theta =0}=0$ due to the TR invariance.
  Since the field strength $B$ is gauge invariant for the gauge transformation $| g ^\prime \rangle = |g \rangle e^{\mi \chi}$,
  $  A ^\prime _\mu  = \langle g ^\prime | \partial _\mu g ^\prime  \rangle
  =A_\mu +\mi \partial _\mu \chi
  $, let us take a temporal gauge by imposing a gauge condition,
  $A^{(t)}_t=0$, which is quite useful for the discussion of the pumping.
\cin{Then} the field strength is given as
 $B = -\remove{\mi }\partial _t A^{(t)}_\theta $.
Now we have a pumped charge between the time period $[t_a,t_b]$ as
\begin{eqnarray*}
 \Delta Q_{[t_a,t_b]}
 &=&  \int_{t_a}^{t_b} dt \, \delta j_x   = \mi  \int_{t_a}^{t_b} dt \, \partial _t A^{(t)}_\theta
 =\mi A^{(t)}_\theta(t)\bigg|^{t_b}_{t_a},  
\end{eqnarray*}
where the integration over $t$ has been carried out but it
needs a special care, as discussed later.
For any (regularly) gauge fixed many body state $| g \rangle   $, 
the state in the temporal gauge is
given by
$
  | g ^{(t)}(\theta,t)\rangle = | g (\theta,t)\rangle e^{\mi \chi(\theta,t)}
$,
  with  the phase factor, $\chi$, that is path dependent and is  explicitly
  given as
  \crep{
$
  \chi (\theta,t)=- {\rm Im\,} \int_C ds_\mu \langle g | \partial _\mu g \rangle   = 
  \mi \int_0^t d \tau\, \langle g (\theta,\tau)
  | \partial _\tau g (\theta,\tau) \rangle
+  \mi \int_0^\theta d \vartheta\, \langle g (\vartheta,0)
  | \partial _\vartheta g (\vartheta,0) \rangle
  $ where the path is  piecewise linear, $C:(0,0)\to (\theta,0 )\to(\theta,t)$.
  }
       {
$
  \chi (\theta,t)=
  \mi \int_0^t d \tau\, A_t(\theta ,\tau )
+  \mi \int_0^\theta d \vartheta\, A_\theta (\vartheta,0)$.
         }
It surely satisfies the gauge condition, 
\crep{
  $  A^{(t)}_t  (\theta,t)= \langle g^{(t)}| \partial _t g^{(t)} \rangle  =0$
}
     {  $  A^{(t)}_t  (\theta,t)= 0$     }
and
one has
\crep{
\begin{eqnarray*}
  A^{(t)}_\theta &&(\theta,t) = \!\langle g| \partial _\theta  g \rangle ^{(t)}
=   
  \langle g (\theta,t)| \partial _\theta g  (\theta,t)\rangle
  \\  &&
  -\partial _\theta 
 \int_0^t d \tau\, \langle g (\theta,\tau)  | \partial _\tau g (\theta,\tau) \rangle
-\langle g (\theta,0)  | \partial _\theta g (\theta,0) \rangle.
\end{eqnarray*}
}
     {
\begin{eqnarray*}
  A^{(t)}_\theta (\theta,t) &=& 
A_\theta (\theta,t)
  -\partial _\theta 
 \int_0^t d \tau\, A_t (\theta,\tau)   
- A_\theta (\theta,0).
\end{eqnarray*}
       }
This is
gauge invariant:  it is directly confirmed but  is clear in a discretized form \cite{Fukui05}
as shown in Fig.\ref{fig:t-gauge}.
Substituting this into $\Delta Q$ above, we have  a  pumped charge  in
a {\it novel} gauge invariant form.
The discussion up to this point is general and applicable  for both 
with/without edges.

\begin{figure}[h]
\includegraphics[width=80mm]{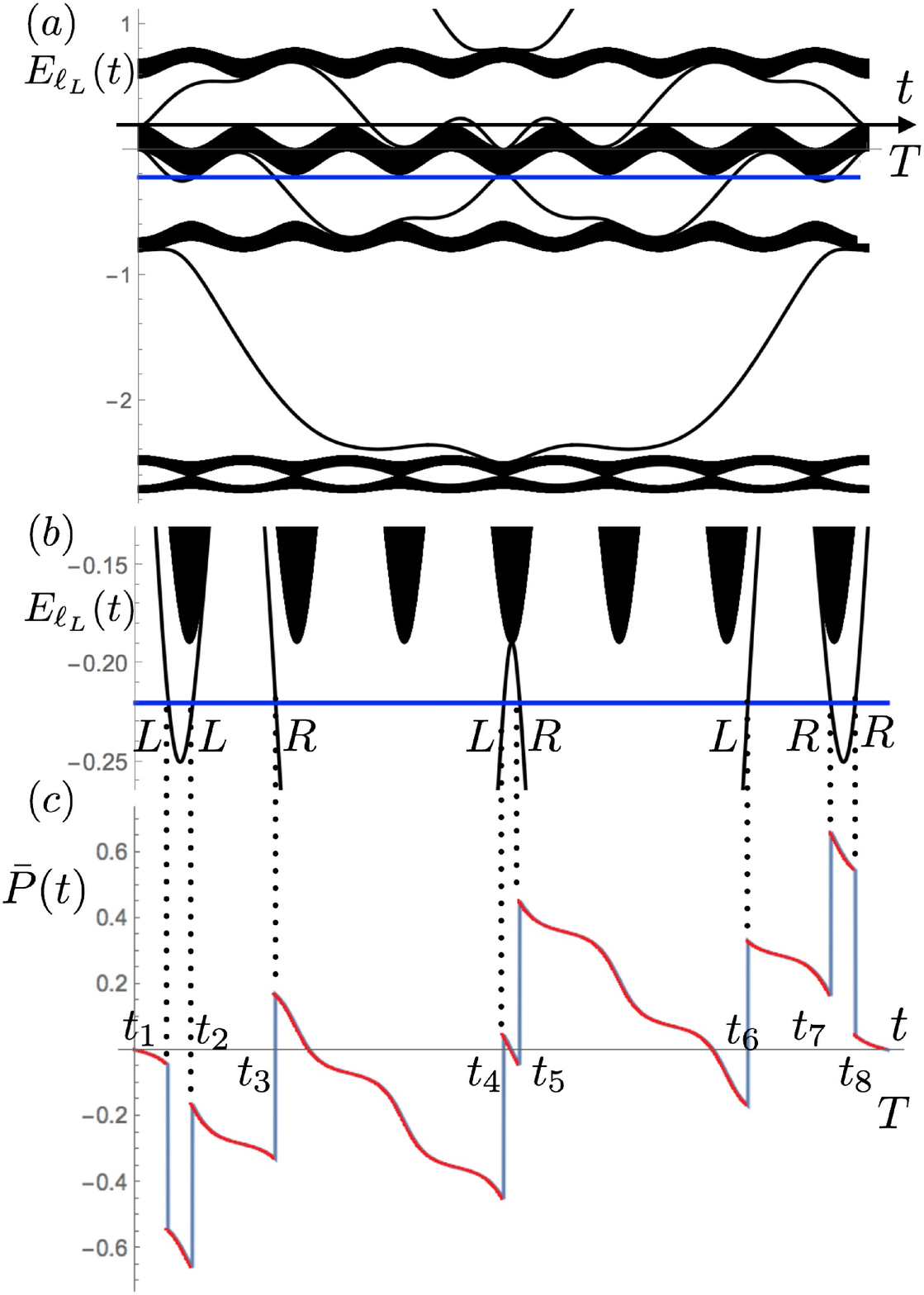}
\caption{\label{fig:example} 
  (a)  One particle spectrum $E_{\ell_L}(t)$
  \cout{of the hamiltonian (black lines)} and the
  fermi energy (blue line)\cout{.}
{:} $t_x=t_y=1$, $\phi=2/7$ and $L_x=1750$ 
{(350)}.
  (b) Enlarged spectrum near the fermi energy.
  (c) The \coutn{normalized} CM, \crep{$\bar P(t)$}{$P(t)$}, with edges by numerical calculation.
\cout{Numerically obtained discontinuities are}$\Delta \crep{\bar P}{P}(t_i)=
-0.499503 (-0.4954)$,
$0.496542 (0.487094)$,
$0.498412 (0.494115)$,
$0.497435 (0.491092)$,
$0.497435 (0.491092)$,
$0.498412 (0.494115)$,
$0.496542 (0.487094)$,
 and 
$-0.499503 (-0.4954)$
   for $i=1,\cdots,8$.
\cout{The total discontinuity is} $\sum_i \Delta \crep{\bar P}{P}(t_i)=1.98577(1.9538)$.
\cout{The results for the system $L_x=350$ are shown in the parentheses.}
}
    \end{figure}

Now let us consider a system with edges 
by imposing the open boundary condition.
In this case, the twist of the hopping $e^{-\mi \theta/L_x}$ is gauged out by the many body gauge transformation 
\begin{eqnarray*}
  {\cal U}(\theta) &=& \prod_{j=1}  e^{-\mi \theta n_j (j-j_0)/L_x},
  \quad j_0=L_x/2,
\end{eqnarray*}
which operates for the fermion operator as $  {\cal U} c_j {\cal U}  ^\dagger = e^{\mi \theta j/L_x}c_j$ and
we have
$| g(\theta )\rangle = {\cal U}|g_0 \rangle  $
where  $|g_0 \rangle $ is a snap shot ground state of the hamiltonian $H(0,t)$.
Then noting that $A_t=\langle g| \partial _t g \rangle
=\langle g_0| \partial _t g_0 \rangle $ is $\theta  $ independent
and
$
  A_\theta = 
  \langle g_0|{\cal U}^\dagger   \partial _\theta {\cal U} | g_0 \rangle
$, 
  the Berry connection {\it in the temporal gauge }
  for the system with edges is
  \crep
  {
\begin{eqnarray*}
  A^{(t),\ed}_\theta 
  &=& -\mi \Delta   P(t),\ 
  P(t)  = 
 \sum_j \frac {j-j_0}{L_x}  \langle g_0(t)|  n_j  | g_0 (t)\rangle,
  \label{eq:CM}
\end{eqnarray*}
where {$P(t)$} is a \coutn{normalized} CM of $|g_0(t) \rangle $
and $\Delta \bar P=\bar P(t)-\bar P(0)$
($-1/2\le \bar P\le 1/2$). 
  }
  {
\begin{eqnarray*}
  A^{(t),\ed}_\theta 
  &=& -\mi \big[  P(t)-P(0)\big],\ 
   P(t)  = \sum\nolimits_j x_j \rho_j(t)
  \label{eq:CM}
\end{eqnarray*}
where
$  \rho_j(t) = \langle g_0(t)|  n_j  | g_0 (t)\rangle$,
$ x_j =  \frac {j-j_0}{L_x}  $  and  $P(t) $ is the CM.
Now we have the pumped charge as
\begin{eqnarray*}
 \Delta Q_{[t_a,t_b]} &=& P(t_b)-P(t_a).
\end{eqnarray*}
Note that this is only well defined for a system with boundaries.
  }
  \crep{It should be  noted }
  {We also stress } that the \coutn{normalized} CM derived here is distinct from the Zak phase for infinite systems.
  We assume
  \crep{$\bar P(t)$}{$P(t)$} is a CM measured 
      for the {\it snap shot ground state }
      {\it in contact with a particle reservoir},  that is, the system
        is specified by the chemical potential $\mu  $ and the
        temperature is sufficiently low.
\cout{        \footnote
            {{\color{blue}
                We expect
                the discussion here is justified even for a closed
                system
                in the limit $L_x\to \infty$,
                since the total number of particles is 
                conserved after the pumping of one cycle.}
            }.            }
It should be  also 
  distinguished from the CM of {\it the time dependent wave function
    \crep{
      $
\sum_j \frac {j-j_0}{L_x}  \langle G(t)|  n_j  | G (t)\rangle,
$
    }
         {
      $
\sum_j x_j  \langle G(t)|  n_j  | G (t)\rangle,
$         }
which is recently observed in real
experiments}
\cite{Nakajima15,Lohse15,Wang:2013fk_pump}.
Even  though the fermi energy is in the bulk gap,
when the one particle energy of the edge state coincides
to the fermi energy, 
{\it the many body   gap} of the snap shot hamiltonian necessarily closes
and the edge state becomes suddenly occupied/unoccupied
(See Fig.\ref{fig:example}).
This sudden change of the snap shot ground state
causes singularities (discontinuities) in $\crep{\bar P}{P}$ since the
edge state is spatially localized and its contribution to
the normalized CM is
$\pm 1/2 $ in the limit $L_x\to \infty$.
This is inevitable since topologically non trivial ground state
is associated with the edge states passing through the gap.
      Then labeling the gap closing time's by $t_i$'s ($t_i<t_{i+1}$, $i=1,2,\cdots$) and
        dividing
      the period into the time intervals without the singularities
      (patch-working in the the time domain),
      the total pumped charge in a single period $T$
      is given as
      \crep{
        \begin{eqnarray*}
\Delta {Q^\ed} 
&=& \sum_{i} \int_{t_i^+}^{t_{i+1}^-} dt\, \partial _t \bar P(t)
=\sum_{i}  \big[\bar P(t_{i+1}^-)-\bar P(t_{i}^+)\big]
\\
&=&
-\sum_{i}  \big[ \bar P(t_{i}^+)-\bar P(t_{i}^-)\big]
= -\sum_i \Delta \bar P(t_i),
        \end{eqnarray*}
      }
      {
    \begin{eqnarray*}
\Delta {Q^\ed} 
&=& \sum_{i} \int_{t_i^+}^{t_{i+1}^-} dt\, \partial _t P(t)
=\sum_{i}  \big[P(t_{i+1}^-)-P(t_{i}^+)\big]
\\
&=&
-\sum_{i}  \big[ P(t_{i}^+)-P(t_{i}^-)\big]
= -\sum_i \Delta P(t_i),
    \end{eqnarray*}
    }
    where $t_i^\pm\equiv t_i\pm 0$ and 
    $\Delta {P}(t_i)\equiv P(t_i^+)-P(t_i^-)$ is a discontinuity of the CM at $t_i$.
    Also we used the periodicity of the CM in the adiabatic limit
    $P(t+T)=P(t)$, which follows from the periodicity of the hamiltonian
    $H(t+T)=H(t)$. 
    This simple expression can be understood as a bulk-edge correspondence
    in time domain.
    We put ''$\ed$'' to mark the expression for the system with edges.
Note here that {\it $P$ is only well defined with edges} and can not be
defined for a system with periodic boundary condition.
Depending on the position of the edge state that cuts the fermi energy and the way it does, this discontinuity is determined as
      \begin{eqnarray*}
        \Delta P(t_i) &=&
\left\{
\begin{array}{ccc}
-  1/2 & \text{Right}:& \text{become\ unoccupied}
  \\
+  1/2 & \text{Right}:& \text{become\ occupied}
  \\
+  1/2 & \text{Left}:& \text{become\ unoccupied}
  \\
-  1/2 & \text{Left}:& \text{become\ occupied},
\end{array}
\right.
      \end{eqnarray*}
      where ``Right'' implies the edge state is localized near the boundary $j=L_x$ and the ``Left'' is for the edge states near the boundary $j\sim 0$.

      Since the pair of the discontinuity coincides to the winding of the
      corresponding edge state energy around the ``hole'' (that corresponds to
      the energy gap) on the Riemann surface \cite{Hatsugai93-edge},
      total discontinuity is given by the winding number
      $I_M$ of the edge states
      \begin{eqnarray*}
\sum_i \Delta P(t_i) &=& -I_M,\qquad      \Delta {Q^\ed}  = I_M,
    \end{eqnarray*}
      where we assume that the fermi energy is in the $M$-th energy gap
      from below. This algebraic definition of the winding number is
      only possible for the Harper equation, that is, only for special
      form of the $v_j(t)$.
      However 
      the relation  is generically justified
      by defining the winding number $I_M $ 
      as the number of paired  edge states 
      with suitable sign depending on the direction of the crossing of
      the spectral flow with the fermi energy.
      This is a modified Laughlin argument \cite{Laughlin81}
      which is widely used for 
      the
      topological number of edge states for
      various topological phases\cite{Hatsugai93-edge,KM05,Qi06}.
      \cin{
        Since the total number of particles is conserved,
the        gapless times 
$t_i$'s that correspond to the
(dis)appearance of the edge state
        are paired (irrespective to the position).
        It guarantees the quantization of the total pumped charge
        $\Delta   Q^\ed$ as an integer since even number of additions
        of $\pm 1/2$ is an integer.
        It is a consequence of the local
        $U(1)$ gauge symmetry as the Laughlin argument,
        \cinf{but has sharp contrast to the QHE in which
          \sout{the discontinuities $\Delta P$ }
               {
                 contribution of the edge states to the Hall conductance}
          are always integers  in the process of a unit flux penetration, since the right/left edge states are always paired on the Fermi surface.}
        This $\pm 1/2$ contribution can be understood as a {\em fractionalization}
        of electrons into massive Dirac fermions\cite{Niemi,spec}.

      }

      Also
counting the topological number \coutn{by the spectral flow}
      with discontinuities here should be compared with the counting of the
       singularities of the $\eta$-invariant for the Atiyah-Patodi-Singer index
       theorem \cite{PSP:2128256,Alvarez85}.
       \cin{
         Here we have clarified the close inter-relation between
         the topological nature of the discontinuities
         and local $U(1)$ gauge symmetry. This is not just a math but \sout{is also experimentally proved recently}
          plays a crucial role in the recent experiments\cite{Nakajima15,Lohse15}. \remove{It is a surprise.}}

      As an example,
      we calculated the \coutn{normalized} CM, $P(t)$,
      numerically for
      $\phi=2/7$ and $L_x=1750$ and $L_x=350$.
      By the clear finite size effects, the discontinuities deviates from
      $\pm 1/2$ but approaches to the quantized values $\pm 1/2$ by the limit
        $L_x\to\infty$.
      In this  case, we have $\Delta Q^\ed=-2$.
      
    \noindent

    Note that although the total  pumped charge is governed by the
    discontinuity $\sum_i\Delta \crep{\bar P}{P}(t_i) $
    due to the edge states,
    pumped charge is not carried by singularities caused by
    the edge states.
    The charge is still carried 
    by the bulk 
    as we explain below.
    This is the bulk-edge correspondence in the topological pumping. 
    As one can see in Fig.\ref{fig:example},
    the  charge is pumped in the intervals between the singularities
    (red lines),
which
    is the bulk contribution.
    Even though the system has boundaries, the effects of the edges
    are negligible for the bulk state since the one particle state of
    the bulk is extended and the amplitude near the boundaries is vanishing in the limit $L_x\to \infty$.

    Although the \coutn{normalized} CM is ill-defined for bulk (both for
      a periodic/infinite system),
    the pumped charge is well-defined as discussed by Thouless　\cite{Thouless83}.
    As for an infinite system, the one particle state is given by the Bloch state
    \cout{as}
    $\psi_{j,\ell_L}\propto e^{\mi k_x j}u_j(k_x)$
{.}\cout{,
    $u_{j+q}(k_x)=u_{j}(k_x) $, 
    where $u_j(k_x)=u_{j+q}(k_x)$ satisfies a reduced $q$ site problem
    replacing the twist $\theta /L_x$ by $ k_x$.
    The translational invariance $\psi_{j+q}(k_x)=e^{\mi k_x q}\psi_j(k_x)$ implies
    $k_x\in[0,\Delta k]$, $\Delta k=2\pi/q$.
    The one particle states for the large periodic system (bulk) $L_x=q l$ ($l\gg 1$) are all reconstructed from  the Bloch state as
$      \psi_{j,\ell}^n 
      =
      e^{\mi \frac { n\Delta k}l \cdot j}u_{j,\ell}(k_x)/\sqrt{l}
$ where
$    k_x 
    = \frac {\Delta k}{l}(n+ \frac {\theta}{2\pi})\in[\frac {n\Delta k}l,\frac {(n+1)\Delta k}{l}) ]$,
($    n = 0,1,\cdots, l-1$) and 
$   \ell = 1,\cdots, q$
 is a band index. }
    \cout{
      Then the many body ground state of the snap shot hamiltonian when the fermi energy is in the $M$-th gap  is given by
$
  | g \rangle = \prod _{\ell=1}^M\prod _{n=0}^{l-1}
( \sum_{j=1}^{L_x} \psi_{j,\ell}^n  c_j ^\dagger )|0 \rangle
  $ and the Berry connection for the $\theta $ direction is given by that of the $U(1)$ part of
  the Berry connection of the filled fermi sea.
    }

        {
      Now let $|g^{(t)} \rangle  $ be a many body ground state of the snap shot hamiltonian in the temporal gauge.
          }
\cout{  By using  the temporal gauge of the Bloch state,
  one} 
{Assuming the fermi energy is in the $M$-th gap,  one} has
  \begin{eqnarray*}
  A^{(t),\bl}_{\theta } &=&   \langle g^{(t)} | \partial_\theta  g^{(t)} \rangle
  =  \int_0^{\Delta k} \frac {dk_x}{2\pi}    \, a^{(t)}_{k_x},
  \\
  a^{(t)}_{k_x } &=&     {\rm Tr} _M {\cal A}^{(t)} _{k_x },\
  {\cal A}^{(t)} _{k_x } = u ^\dagger \partial_{k_x } u,\ u= (\bm{u} _1,\cdots,\bm{u} _M),
  \end{eqnarray*}
  where 
  the limit $L_x\to\infty$ is taken and additional gauge condition
  $a^{(t)}_t=   {\rm Tr} _M u^\dagger   \partial_t  u  =0$ is imposed
{.}
\cout{  for the $U(1)$ part of the Bloch states $u$.}
Also we put ``$\bl$'' to specify that it is purely from bulk.
  By using the gauge invariant form of the temporal gauge,  
    \begin{eqnarray*}
      a^{(t)}_{k_x} (k_x,t)&=& a_{k_x}(k_x,t)- \partial _{k_x}\int_0^t d\tau\, a_t(k_x,\tau) - a_{k_x}(k_x,0),
    \end{eqnarray*}
    the time derivative
    of $A^{(t),\bl}_\theta $ is written by
    the field strength of the Bloch state,
    $b=\partial _{k_x} a_t-\partial _t a_{k_x} =-\partial _t a^{(t)}_{k_x}$
{.}
    \cout{
, as      
    \begin{eqnarray*}
\partial _t  A^{(t),\bl}_{\theta } =
  \int_0^{\Delta k} \frac {dk_x}{2\pi}
\partial _t   a^{(t)}_{k_x} (k_x,t)
= 
-\int_0^{\Delta k} \frac {dk_x}{2\pi}
b(k_x,t).
    \end{eqnarray*}
    }
    Now 
    the bulk contribution of the pumped charge
    between the time period $[t_a,t_b]$ is written as\cite{Thouless83}
\begin{eqnarray*} 
  \Delta Q^\bl _{[t_a,t_b]} &=& 
  \mi\int_{t_a}^{t_b}dt \partial _t  A^{(t),\bl}_{\theta }
=\frac{1}{2\pi\mi}  \int_{t_a}^{t_b} dt  \int_0^{\Delta k}  {dk_x}  b(k_x,t).
\end{eqnarray*}
\cout{These equations imply}
{It implies}
that \cout{the average of the
  temporal gauge of the Bloch function $\mi/(2\pi) \int_0^{\Delta k}dk_x  \,  a^{(t)}_{k_x}(k_x,t)$}
{$\mi\partial _t  A^{(t),\bl}_{\theta } $}
is an {\it effective CM} 
\cout{at
  the time $t$} 
{of the bulk}, even though the CM itself
is not well defined\cout{for the bulk}.

\begin{figure}[h]
\includegraphics[width=70mm]{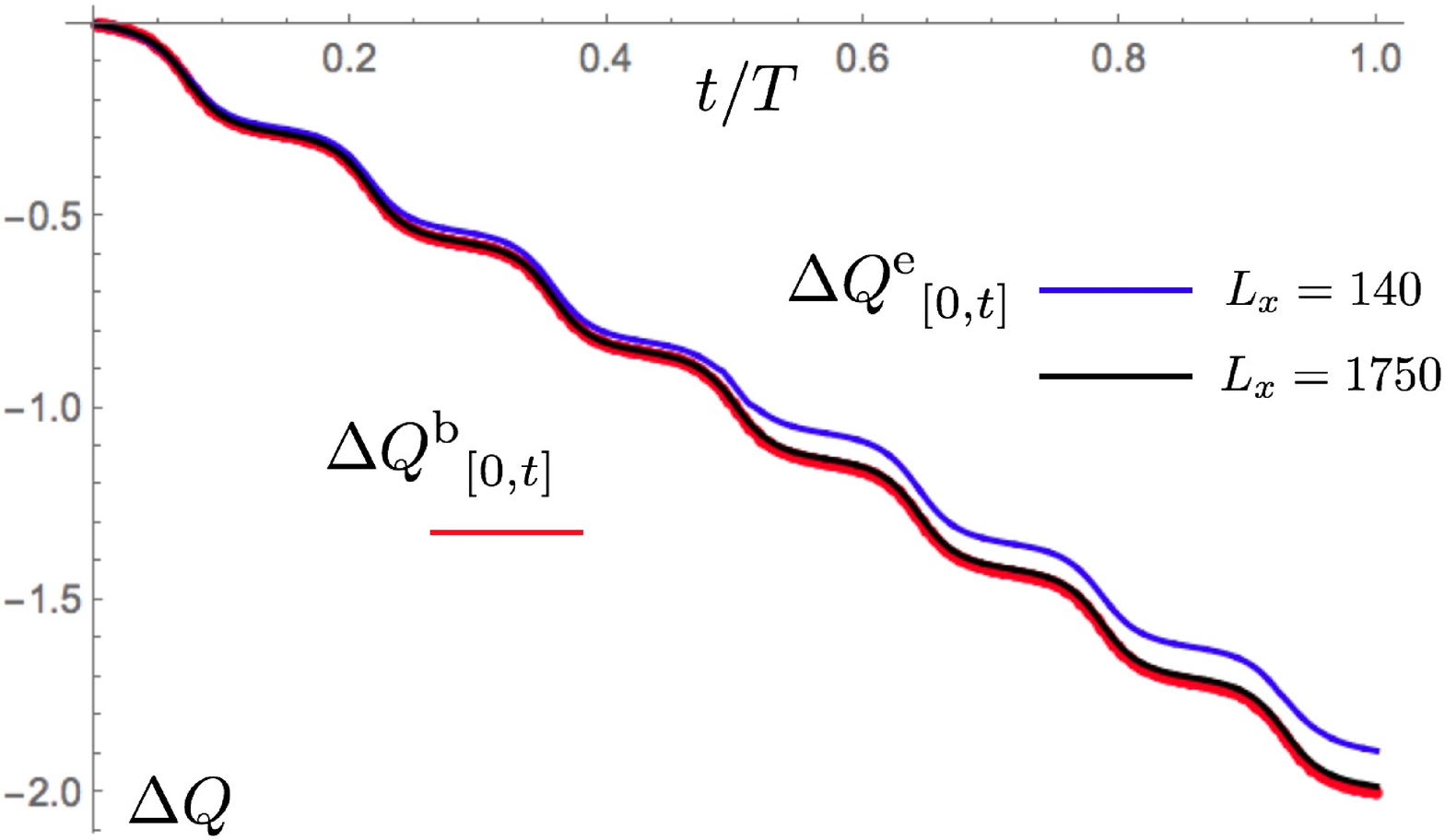}
\caption{
  \label{fig:bulk-edge-pump} 
  \cout{Total amount of the pumped} 
{Pumped} charge between the time interval $[0,t]$.
  The red line is by the bulk (Bloch states)
  $\Delta Q^\bl_{[0,t]}$. 
  The shifted pumped charge $\Delta {Q^\ed} _{[0,t]}$'s are shown \cout{by} 
{for} $L_x=140$ (blue)
  and $L_x=1750$ (black) lines.
  The parameters are the same as Fig.\ref{fig:example}.
}
\end{figure}

To demonstrate
the contribution of the pumping with edges
is from  the bulk,
let us define a pumped charge between the time period $[t_a,t_b]$
by compensating (skipping) the singularities from the edges as
    \begin{eqnarray*}
      \Delta Q^\ed  _{[t_a,t_b]} 
      &=&
     \crep{ \bar  P}{P}(t_b)-\crep{\bar P}{P}(t_a)
      -
        \int_{t_a}^{t_b} d\tau\,
        \sum_i \Delta \crep{\bar P}{P}(t_i) \delta (\tau-t_i).
      \label{eq:pumped-e-period}
    \end{eqnarray*}
      This is a bulk contribution to the \coutn{normalized} CM for
      the system with edges.
    According to the consideration here, 
      we  expect this bulk contribution
      $\Delta Q^\ed  _{[t_a,t_b]} $  asymptotically approaches to
$\Delta Q^\bl _{[t_a,t_b]}$ in the large system.
    We have numerically   evaluated them 
    for $\phi=2/7$  and shown in Fig.\ref{fig:bulk-edge-pump}. They show clear coincidence within numerical errors, indicating the relation between the bulk and the edges.

{The}\cout{Further the
 total} pumped charge by bulk in a cycle
is given by the Chern number $C_M$ \cite{Thouless83}
of the Berry connection $a_\mu $ 
as $\Delta Q^\bl=  \Delta Q^\bl _{[0,T]}= C_M $.
\cout{
  If we interpret $b$ as
$b = \partial _{k_x} a_{k_y}-\partial _{k_y}a_{k_x}$ ($\partial _{k_y}\equiv \partial _t$, $a_{k_y}\equiv a_t$ ),
it is a natural definition of the Chern number in  QHE,
which corresponds to the Chern number of the filled $M$ bands (Bloch states) in the present pumping.
}
Now we have established the physical bulk-edge correspondence
in the topological pumping
as
\begin{eqnarray*}
  \Delta Q^\ed       &=&     \Delta Q^\bl      
\end{eqnarray*}
{that} gives the relation between the two topological invariants as was
also shown in the QHE
as
\begin{eqnarray*}
I_M &=&   C_M .
\end{eqnarray*}
It guarantees quantization of the Chern number $C_M$, since
$I_M$ due to the singularities of edge states
is quantized by definition.



{
  \cin{
    Generically the bulk-edge correspondence is for the
    manybody state and the bulk and the edge can not be separated.
    Without gapped bulk, edge states are not well defined.
    We put a stress on
    that in the present pumping of the non interacting system, the manybody state
    is constructed from the one particle states which are classified into
    the bulk(unnormalizable) and the edge(localized). Then the contributions to
    the pumped charge are also clearly separated (Fig.\ref{fig:bulk-edge-pump}).
  }

{
    To describe the topological  pumping by the CM,
    one necessarily needs to use a system with edges
    in contact with a particle reservoir.
      As we have established here,
      the pumped charge of the system with edges
      is carried by the bulk not by the edge states
          and the CM observed in the cold atom experiments
          \crep{    $P_{{\rm exp} }(t)=\langle G(t)|\sum_j n_j\frac {j-j_0}{L_x} |G(t) \rangle $}
{$P_{{\rm exp} }(t)=\sum_j x_j\langle G(t)| n_j|G(t) \rangle $}          
    is described by the bulk.
  Still the quantization of the pumped charge is governed by the edge states where
    the edge states induce discontinuities in the \coutn{normalized} CM 
    in the adiabatic limit.
      Even though the \coutn{normalized} CM shows singularities in the adiabatic limit,
    {\it
      these singularities can not be observed in realistic experiments
      in cold atoms}, since the appearance of the edge states at the fermi energy
      necessarily implies the breakdown of the adiabaticity.
      As in an  realistic experiment with
      finite speed pumping, 
      the many body wave function
      approximately remains unchanged passing through the
      gapless point,
      that is, the bulk is 
      adiabatic but the edge states
      is well described by the sudden approximation.
      It causes excitation across the gap, which eventually results in
      accumulation/depletion of the surface charges after one cycle.
      These two conditions, as the bulk is adiabatic and the edge is sudden,
      are  conflicting requirements, if one requires both limits rigorously.
      Then in the  experiment,
      it may not be easy to observe exact quantization of
      the pumped charge, especially for cases with high Chern numbers.
      We have also confirmed consistency of the 
      argument
      by a direct numerical
    integration of the  Liouville$-$von Neumann equation\cout{,
      $\mi \hbar    \partial _t \rho(t)=\cmt{H(t)}{\rho(t)}$,}
      for
      the density matrix
{.}\cout{ $\rho(t)$.} 

    %
    
{\color{green}\sout{
  To summarize,
      the topological pumping, proposed by Thouless 
      and is observed by
measuring the CM  of the system with edges, is governed
by the edge states.
The bulk is hidden but carries the pumped charge in the experiments
with edges.
The discontinuities of the \coutn{normalized} CM in the adiabatic limit
associate with the edge states 
can  not be measured experimentally
but guarantees the quantization of the pumped charge 
as the bulk-edge correspondence in the topological pumping.
}}

We thank stimulating discussions on the topological pump with Y. Takahashi, S. Nakajima and M. Lohse.
The work is supported in part by
KAKENHI from JSPS 
(Nos.26247064, 25400388,16K13845).
%

\end{document}